\documentstyle[preprint,aps,epsf,floats]{revtex}
\renewcommand{\baselinestretch}{1.15} 

\newcommand{\r}{{\bf r}}

\newcommand{\ep}{\epsilon}

\begin{document}
\title{\large 
{\bf Thermodynamics and transport \\ in mesoscopic disordered networks}}
 \author{M. Pascaud and G.
Montambaux} \address{Laboratoire de Physique des Solides,  associ\'e au
CNRS \\ Universit\'{e} Paris--Sud \\ 91405 Orsay, France}
\maketitle
\widetext

\noindent
 \begin{center}
 \begin{abstract}
\parbox{14cm}{
We describe the effects of  phase coherence on  transport and thermodynamic
properties of a disordered conducting network.
In analogy with weak-localization correction,
we calculate the phase coherence contribution to  the magnetic response of
mesoscopic metallic isolated networks. It is
 related to
the return probability for a diffusive particle on the corresponding network.
By solving the diffusion equation on various types of networks,
including a ring with arms, an infinite square network or a chain of
connected rings, we deduce the magnetic response. As it is the case for
transport properties --weak-localization corrections or universal
conductance fluctuations-- the magnetic response can be written
in term of a single function $S$ called spectral function which is
related to  the {\it spatial average} of the return probability on the
network. We have found that the magnetization of an ensemble of {\bf
connected} rings is of the same order of magnitude as if the rings were
disconnected.} \end{abstract}
\end{center} 

\section{Introduction}

The understanding of persistent currents in mesoscopic rings is still an open
problem since no quantitative agreement has yet been reached  between
theoretical developments and experimental
results\cite{Buttiker83,Levy90,Chandrasekhar91,Mailly93,Reulet95,Webb95,Montambaux95}.
The only quantitative
calculations of persistent currents relevant to the experimental situations
rely on analytical methods in which disorder and interactions are treated
perturbatively
\cite{Cheung89,Ambegaokar90,Schmid91,VOppen91,Altshuler91,Akkermans91,Oh91,Argaman93a,Argaman93b,Montambaux96}.
The disorder is described in the usual framework of perturbation theory
and interactions are treated in a Hartree-Fock approximation.

In ring experiments, the important parameters are certainly the
strength of disorder
and the strength of interactions. Both parameters are difficult to vary
experimentally in a controlled way.
In order to go beyond this problem, we have recently
proposed to extend the analytical method to describe new geometries
which
generalize the ring topology, and which allow the variation of some
geometric parameters in order
to change the amplitude of the persistent current\cite{Pascaud97}. In
the experiment done on a single
Ga-As ring, for example, the ring was connected to leads\cite{Mailly93}
and one
may wonder how the current is changed when the lengths of the leads
increase.
On the other hand, experiments on many rings have been performed on an
ensemble of  {\it disconnected} rings\cite{Levy90,Reulet95}. One may wonder
whether
the same effect persists when the rings are {\it connected}, i.e. what is the
 magnetization of a lattice, resulting from phase coherence.

Such a generalization, from the ring geometry to a network, had
already been considered in the past to describe
the  weak-localization correction to the conductivity.
This correction is related to the return probability,
 solution of a diffusion equation\cite{Khmelnitskii84}. This
return probability  has two components,
 a purely classical one and an interference term which results
from interferences between pairs of time-reversed trajectories. In the
diagrammatic picture, they are related to the diffuson and Cooperon diagrams.
The interference term, $p_\gamma(\r,\r',\omega)$, is field dependent
and  is
solution of the diffusion equation:
\begin{equation}  \label{diff}
[\gamma - i\omega -  D({\bf \nabla} + {2 i e {\bf A}\over \hbar c
})^2]p_\gamma(\r,\r',\omega)=  \delta(\r-\r')  \end{equation}
where $D$ is the diffusion coefficient. The scattering rate $\gamma = D
/L_\phi^2$ describes the breaking of phase coherence.
$L_\phi$ is the phase coherence length. $\gamma$ will be compared to
$1/\tau_D=D/L^2$ where $\tau_D$ is the diffusion time, typical time to
diffuse through the system of size $L$. This time is the inverse of the
Thouless energy. The classical probability obeys the same equation as
(\ref{diff}) with ${\bf
A}=0$. By solving the diffusion equation (\ref{diff}) on various lattices,
the weak-localization correction was calculated by Dou\c cot and
Rammal\cite{Doucot86}.
The persistent current of a ring, or in a more general geometry the
mesoscopic
magnetization, have a different nature. They are thermodynamic and not
transport quantities.
Transport properties exhibit weak-localization effects because they are
response functions related to the propagation of two states: the conductivity is written in terms of a product of two Green functions whose disorder average  is related to the return probability.
A priori, average thermodynamic quantities do not exhibit weak-localization
corrections since they
are integrals of the density of states, i.e. they are written in terms of a
single Green function.
This is why the average persistent current $\langle I \rangle$ of non
interacting electrons vanishes for an ensemble of rings. However, the
typical current, $I_{typ}= \sqrt{\langle I^2 \rangle}$ involves the product
of two propagators. Similarly, the average Hartree-Fock correction
to the average current does not vanish because the interaction term also contains two propagators. As a result, these quantities are expected to be related to weak-localization effects and to the return probability.

This paper is organized as follows: in the next section, we  recall
 how to solve the diffusion equation on a lattice and we calculate the
weak-localization correction.
 In section III, the persistent current and the mesoscopic magnetization are
related to the return probability and are calculated for a few geometries
in section IV. In the last section, we show how  the universal conductance
fluctuations are
related to the return probability and we calculate them for the simple
geometry of the single wire, using this simple approach.

\section{Diffusion on a lattice - Weak-Localization correction}

Let us first  recall that the weak-localization correction to the
conductance of a connected mesocopic sample  can  be written in term
 of the interference part of the return probability
\cite{Khmelnitskii84,Doucot86}:
\begin{equation}\Delta \sigma(r) =-2 s (e^2 / h) D
C_\gamma(\r,\r)\label{WL}\end{equation}
 $s$ is the spin degeneracy. The
Cooperon $C_\gamma(\r,\r,H)$
 is the time integrated field-dependent return probability:
$C_\gamma(\r,\r,H) = \int_0^\infty p_\gamma(\r,\r,t,H)
dt=p_\gamma(\r,\r,\omega=0,H)$.
Dou\c cot and Rammal\cite{Doucot86} have  calculated the
weak-localization corrections
on  various networks.
Considering networks made of quasi-1D wires, so that the diffusion can
be described as one-dimensional, the
Cooperon $C_\gamma(r,r')$ obeys the one-dimensional diffusion equation
\begin{equation}   \label{diff2}
[\gamma -  D (\nabla + {2 i e A\over \hbar c })^2]
C_\gamma(r,r')=\delta(r-r')
\end{equation}
with the continuity equations written for every node $\alpha$ (including
the starting point $r'$ that can be considered as an additional node in
the lattice)\cite{Doucot86} \begin{equation} \label{cal}
\sum_\beta ( -i {\partial \over \partial r} + {2 e A \over \hbar c})
C_\gamma(r,r')|_{r=\alpha} = {i \over D S }
\delta_{r',\alpha}
\end{equation}
$r,r'$ are  linear coordinates on the network. The sum is taken over all
links relating the node $\alpha$ to its neighboring nodes $\beta$.
Integration of the differential
equation (\ref{diff2}) with
the boundary conditions (\ref{cal}) leads to the so-called network equations
which relate $C_\gamma(\alpha,r')$  to the
neighboring  $C_\gamma(\beta,r')$.
\begin{equation} \sum_\beta\coth({l_{\alpha\beta} \over
L_\phi})C(\alpha,r')-
\sum_\beta{C(\beta,r')e^{-i\gamma_{\alpha\beta}}\over\sinh(l_{\alpha\beta}
/L_\phi)}
={L_\phi \over D S} \delta_{\alpha,r'}
\label{network}
\end{equation}
$l_{\alpha\beta}$ is the length of the link $(\alpha\beta)$ and
$\gamma_{\alpha\beta}=(4 \pi /\phi_0)\int_\alpha^\beta {\bf A} d {\bf l}$
 is the circulation of the vector potential along this link.
Solving this set of linear equations and
performing the  spatial
integration of $C_\gamma(r',r')$ give access to the
weak-localization correction.

\section{Persistent currents and mesoscopic magnetization}

The magnetization $M(H)$ is the derivative
of the free energy $F$ with respect to the  magnetic field $H$: $M =
-{\partial F \over \partial H}$.
Introducing the
field dependent DOS (for one spin direction),   $\rho(\ep,H)$,
the magnetization
is written at zero temperature as (taking the spin into account):
\begin{equation} \label{Isum}
M = - 2{\partial \over \partial H} \int_{-\ep_F}^0 \ep \rho(\ep,H)
d\ep \end{equation}
The origin of energies is taken at the Fermi energy.
The average magnetization is thus related to the field dependence of the
average density of states $\langle \rho(\ep,H) \rangle$. In a bulk system,
this leads to the Landau diamagnetism. In the thin ring geometry where no
field penetrates the conductor, the DOS only depends on the
Aharonov-Bohm flux. Its average is flux independent
 because the flux modifies only the phase factors of the propagator
which cancel in average.
During the last years, there has been a large amount of work to explain the
origin of a non-zero average current, as observed experimentally.
The constraint that the number of particles is fixed in each ring leads to
an additional contribution to the average current often refered as the
"canonical" current\cite{Bouchiat89,Imry89}. Its amplitude is however very
small\cite{Schmid91,VOppen91,Altshuler91,Akkermans91}. As we
will see later, a larger  contribution comes from the effect of interactions.
\medskip

We first calculate the
typical
magnetization $M_{typ}$, defined as $M_{typ}^2 =
 \langle M^2 \rangle - \langle M \rangle^2$. From eq.(\ref{Isum}), it can
be written as:  \begin{equation}
 M^2_{typ} = 4 {\partial \over \partial H}
{\partial \over \partial H'} \int_{-\ep_F}^0 \int_{-\ep_F}^0 \ep \ep'
K(\ep-\ep',H,H') d\ep d\ep'  \label{Itypical}
\end{equation}
where $K$ is the correlation function of the DOS: $K(\ep -
\ep',H,H')=\langle \rho(\ep,H)\rho(\ep',H')\rangle - \rho_0^2$.
 $K(\varepsilon)$ has been
calculated by Altshuler
and Shklovskii\cite{Altshuler86} for a bulk system and later in the presence
of a magnetic flux\cite{Schmid91,VOppen91,Altshuler91,Akkermans91}
for the ring geometry. A very useful semiclassical
picture has been presented by Argaman {\it et al.}, which
relates the form factor $\tilde K(t)$, the Fourier transform of
$K(\varepsilon)$,  to the integrated return probability
$P(t) = \int p(\r,\r,t) d\r $
for a diffusive particle\cite{Argaman93a}:
\begin{equation}\tilde K(t) = t P(t) / (4 \pi^2)\label{AIS}\end{equation}

In this picture, like the return probability, the form factor
is the sum of a classical and an interference term:

\begin{equation}  \label{decouplingK}
\tilde K(t,H,H')= {t \over 4 \pi ^2 }[
P_{\gamma}(t,{H-H'\over
2})+ P_{\gamma}(t,{H+H'\over 2})]
 \end{equation}
 Fourier transforming $K(\ep-\ep')$ and using the identity
 $\int_0^{\infty} \ep d \ep e^{i\ep t} =-1/t^2$, one obtains
straightforwardly\footnote{The lower bound of the time integrals is
actually the mean collision time $\tau_e$ above which diffusion takes
place.}: \begin{equation}  \label{Mtyp}
M_{typ}^2(H)={1 \over 2 \pi^2}
\int_0^\infty {P^"_\gamma(t,H)|^H_0 \over t^3} dt \,,
\end{equation}
where $P^"_\gamma(t,H)|^H_0=   \partial^2 P_\gamma/\partial H^2|_H -
\partial^2 P_\gamma/\partial H^2|_0$.
\medskip

We now turn to the average magnetization.
It has been found
 that the
 Hartree-Fock(HF) correction to the energy leads to an average persistent
current in a ring\cite{Ambegaokar90,Schmid91}.
The average
magnetization can be written as\cite{Argaman93b,Montambaux95}:
\begin{equation}  \label{currentee}
 \langle M_{ee} \rangle =  -\langle{\partial E_T \over \partial
H}\rangle = -{U \over
4} {\partial \over \partial H} \int \langle n^2(\r) \rangle
d\r
\end{equation}
where the
the screened Coulomb interaction is
$ U=4 \pi e^2 / q_{TF}^2$\cite{Eckern91}.  $q_{TF}$  is the Thomas-Fermi
vector. The magnetization can be rewritten in terms of the two-point
correlation
function of the local density of states and then in term of the return
probability\cite{Montambaux96}:
 \begin{equation} \langle M_{ee} \rangle =  -{U \rho_0
 \over  \pi} {\partial \over \partial H} \int_0^\infty  {P_\gamma(t,H) \over
t^2}   dt
\label{currentee3}
\end{equation}
\medskip

It appears that $\langle M_{ee} \rangle$ and $\langle M_{typ} \rangle$
 are time integrals
of the return probability with various power-law weighting functions.
Noting that $P_\gamma(t)$ has the form $P_0(t)e^{-\gamma t}$, all these
quantities can be written as integrals of a single function
$S(\gamma,H)$ that
we call the spectral function\footnote{This function $S$ is related to the
logarithm of the spectral determinant defined in ref.\cite{Andreev95}.
The number variance for a closed system can be written as $$\Sigma^2(E)= {2
s^2 \over \beta \pi^2} \Re [S(\gamma) - S(\gamma+i E)]$$ where $s$ is the
spin degeneracy and $\beta=1$ in zero field, $\beta=2$ if time reversal
symmetry is broken.} \begin{equation}
\label{PC}
S(\gamma,H)=\int {P_0(t,H) \over t } e^{-\gamma t} dt =  \int_\gamma^\infty
d\gamma \int   C_\gamma(\r,\r,H) d\r
\end{equation}
The different magnetizations can be given in terms of the successive
integrals of this function:
 \begin{eqnarray}
\label{Mee3}  \langle M_{ee}(H)\rangle &=& -{U\rho_0 \over \pi}{\partial
\over \partial H}  S^{(1)}(\gamma,H)   \\
\label{Mtyp3} M_{typ}^2(H) &=& {1 \over 2 \pi^2}{\partial^2 \over \partial
H^2} S^{(2)}(\gamma,H)  \big|^H_0
\end{eqnarray}
where    $S^{(n)}(\gamma) = \int_\gamma^\infty d\gamma_n ...
\int_{\gamma_2}^\infty d\gamma_1
S(\gamma_1) $.

\section{Examples}


We first consider the case of a ring of perimeter $L$
connected to
one arm of length $b$, see fig.1.
Such a geometry has been considered in the strictly
 1D case without disorder\cite{Buttiker85,Buttiker94,Akkermans91,Mello93}.
 It is expected that since the electrons will spend some
time in the arm where there are not sensitive to the flux, the magnetization
 will be decreased.
From eqs. (\ref{diff2},\ref{network}), the function
$C_\gamma(r,r,H)$ can be straightforwardly calculated
on the arm and on the
ring. Spatial integration gives:
$$
S(\gamma,H)=- \ln[
{1 \over 2} \tanh {b\over L_\phi} \sinh{L\over L_\phi}+ \cosh
{L\over L_\phi} - \cos 4 \pi \varphi ]
$$
where $\varphi = \Phi/\phi_0$. $\Phi$  is the flux through the ring,
$\phi_0 =h/e$ is the flux quantum. $\gamma \tau_D= (L / L_\phi)^2$.
$\langle M_{ee} \rangle$ and $M_{typ}$ are given by successive integrations
over $\gamma$ according to eqs.(\ref{Mee3},\ref{Mtyp3}).
They can be compared to the case of a single loop without arm.
We find that  they decrease
when $b$ increases and  saturate to a finite value when $b \gg L_\phi$.
The
$m^{th}$ harmonics of the flux dependence of $\langle M_{ee} \rangle$ is
smaller by a factor  $(2/3)^m$ and the
$m^{th}$ harmonics of $ M_{typ} $ is
smaller by a factor  $(2/3)^{m/2}$.

In the case of a ring connected to two diametrically
opposite infinitely long arms, fig.1,
 the m$^{th}$ harmonics of $\langle M_{ee} \rangle$
 is smaller by a factor $(4/9)^m$ and  the reduction factor is
$(2/3)^m$  for the
typical magnetization.
This  result is relevant for the experiment of ref.\cite{Mailly93} where the
magnetization is measured for open and connected rings. Moreover, in the
limit $b \gg L_\phi$, the magnetization
 should be unchanged if reservoirs are attached to the
arms.
We propose that single ring experiments with appropriately designed
arms should be able to measure these reductions.
\medskip

We now turn to the case of an infinite square lattice
whose
average magnetization will be compared with the one of an array of
isolated rings.  The eigenvalues of the diffusion equation can
be calculated for a rational flux per plaquette $\varphi=H a^2/\phi_0=p/2q$.
$a$ is the lattice parameter. Defining $\eta = a/L_\phi$, we
find that the spectral function $S(\gamma,H)$
 is \begin{equation}  \label{MNnetwork}
S(\gamma,H)=
- {1 \over q}\sum_{i=1}^q \langle\langle \ln(4 \cosh \eta
-\varepsilon_i(\theta,\mu))\rangle\rangle \end{equation}
where
$\langle\langle (\ldots) \rangle\rangle = \int_0^{2\pi} {d\theta  \over 2
\pi}
\int_0^{2\pi} {d\mu \over 2 \pi} (\ldots) $. \ $\varepsilon_i(\theta,\mu)$
are the solutions of the determinental equation
$det M = 0$ where the $q \times q$ matrix $M$ is defined by
 $M_{nn} = 2 \cos(4 n\pi \varphi+\theta/q) - \varepsilon$,
 $M_{n,n+1}= M_{n+1,n}= 1$ for $n \leq q-1$ and $M_{1,q}=M^*_{q,1}= \exp(i
\mu)$. This is the matrix
associated to the Harper equation known to be also relevant for other related
problems like tight-binding electrons in a magnetic field\cite{Hofstadter76}
or superconducting networks in a field\cite{Rammal83}.

The expression (\ref{MNnetwork}) can be compared to the
spectral function for a square
ring of perimeter $L=4 a$:
\begin{equation}  \label{MNring}
S(\gamma,H) =
-\ln( \cosh 4 \eta -\cos 4 \pi \varphi)
\end{equation}
From eq.(\ref{Mee3}), one calculates   $\langle M_{ee} \rangle$:
$$
\langle M_{ee} \rangle = U \rho_0{e D \over \pi^2 q} {\partial
\over \partial \varphi } \sum_{i=1}^q \int_\eta^\infty \{ \ln(4 \cosh \eta
-\varepsilon_i(\theta,\mu))\} \eta d \eta 
$$which is a continuous and derivable function of the field. It  can be
compared with  the ring magnetization which can be cast in the form:
\begin{equation}  \label{Meering}
\langle M_{ee} \rangle =  U \rho_0{4 e D \over \pi }
\int_\eta^\infty {\sin 4 \pi \varphi \over  \cosh 4 \eta -\cos 4 \pi
\varphi} \eta d\eta  \end{equation}


The magnetization density is plotted on fig.2 for the ring and the
infinite lattice.
It is seen that the network magnetization density
is about 25  times smaller than the
ring magnetization. 
Considering that on the array of square rings already considered
experimentally\cite{Levy90,Reulet95}, the distance between rings is
equal to the size of the squares,
the number of squares is four times larger when they are connected.
One then expects only a factor of order $6$ between the
magnetization of the array of disconnected rings and the lattice.

We have also calculated the magnetization of a chain of rings connected
with arms of similar length, an obvious
generalization of the experiment done in ref.\cite{Webb95} and we find that
when the rings are connected the average Hartree-Fock magnetization is
reduced by a factor $3$.
\medskip

In conclusion, we showed that persistent current experiments do not
necessarily require the rings to be disconnected and we suggest to perform
magnetization experiments  on arrays of connected rings.

\section{Transport}
We now come back to the calculation of transport properties in term of the
spectral function $S(\gamma)$, with application to the well-known case of an
open wire as an example. From eq.(\ref{WL}), the dimensionless conductance
$g=G/(e^2/h)$ is given by (for one spin direction):

$$\langle \delta g  \rangle = - 2
\int P(t) {d t \over \tau_D}$$
where $\tau_D=L^2/D$ is the diffusion time, inverse of the Thouless energy.
This correction is easily written as a function of $S(\gamma)$:
$$\langle \delta g  \rangle = 2  {1 \over \tau_D}
{\partial \over \partial \gamma} = 2   {\partial S \over \partial x}$$
where  $x= \gamma \tau_D= (L / L_\varphi)^2$.
\medskip

Similarly the conductance fluctuation can be found directly as a function of
the diffusion probability $p(r,r',t)$\cite{Argaman96}. Then, using some
properties of the diffusion propagator, it can be cast in the form, for one
spin direction:    $$\langle \delta
g^2 \rangle = 12
 \int  P(t){t dt \over \tau_D^2}$$

	     or

$$\langle \delta g^2 \rangle = 12 {1 \over \tau_D^2}
{\partial^2 S\over \partial \gamma^2} = 12
 {\partial^2 S \over \partial x^2}$$

In the example of an open wire, the spectral function $S$ is found to
be, considering the appropriate boundary conditions:
$$S(x) = -\ln {\sinh \sqrt{x} \over \sqrt{x}}   $$

from which the weak-localization correction and the variance of the
conductance fluctuations is immediately found.  In particular, in the limit
of complete phase coherence $L_\varphi \gg L$, i.e. $x \rightarrow 0$,
the expansion of the function $S(x)$: $$S(x)\rightarrow - {x \over 6} + {x^2
\over 180}$$ immediately leads to the known universal values:

$$\langle \delta g \rangle = 2  S'(0) = -1/3$$
$$\langle \delta g^2\rangle =12  S''(0)= 2  /15 $$
The variance should be multiplied by two when there is time reversal
symmetry. When $L_\phi$ is finite, the variance of the fluctuations is
simply related to the mean weak-localization correction:
\begin{equation} \langle \delta g^2\rangle =-6 {\partial \over
\partial x} \langle \delta g \rangle \end{equation}
\medskip

In conclusion, we have written various thermodynamic and transport
properties in term of a single function called spectral function, which is
related to the spatial average of the return probability.
We are thus able to calculate these physical properties for any type of
network.

\bigskip

Part of the work was done at the Institute of Theoretical
Physics, UCSB,  and supported by
the
NSF Grant No. PHY94-07194.

\renewcommand{\baselinestretch}{1} 

\newpage
 \begin{figure}[hb]
\centerline{
\epsfxsize  8cm
\epsffile{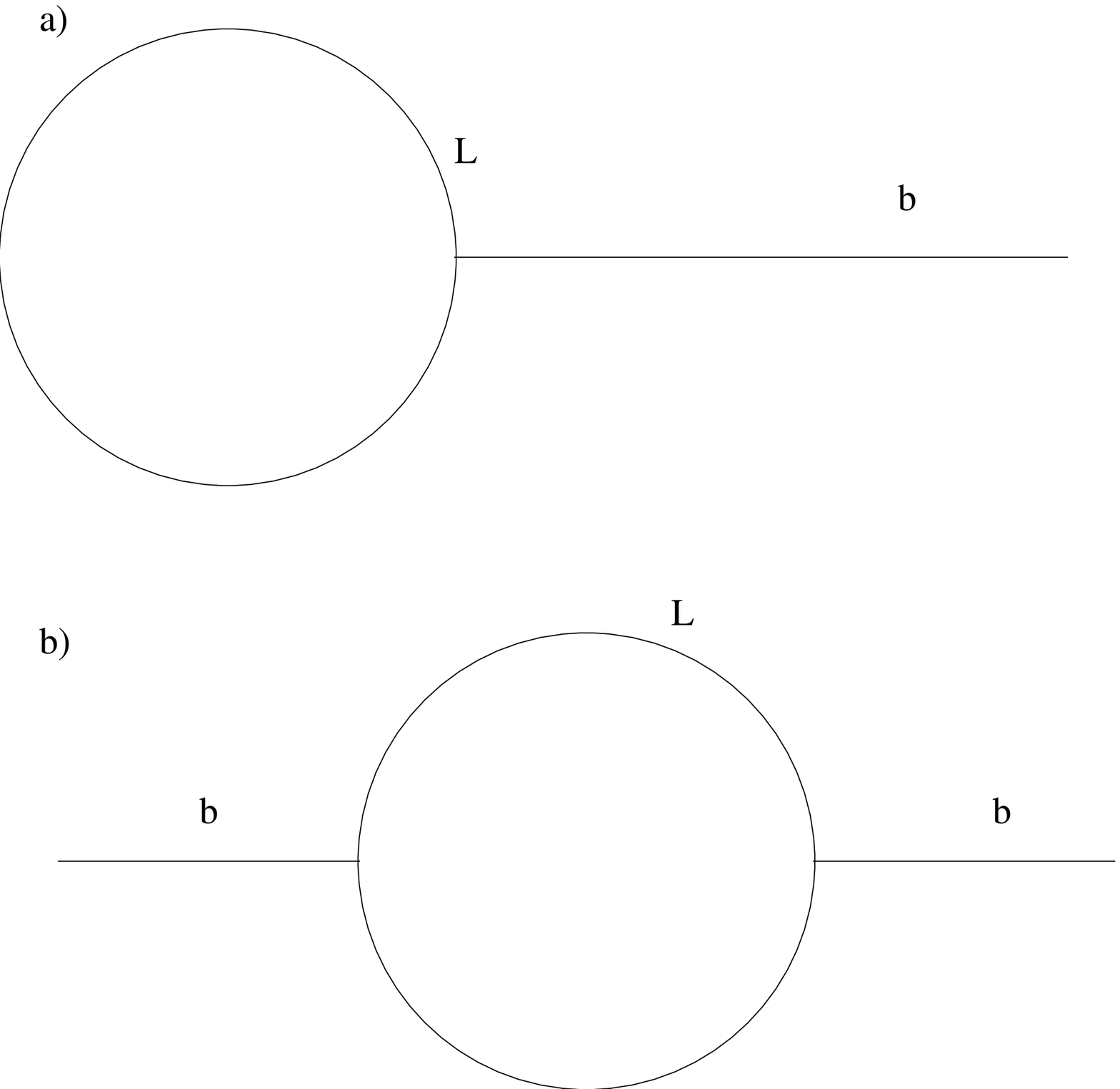}}
\caption{Geometries of a ring with arms considered in the text
} \label{fig1}
\end{figure}

 \begin{figure}
\centerline{
\epsfxsize  10cm
\epsffile{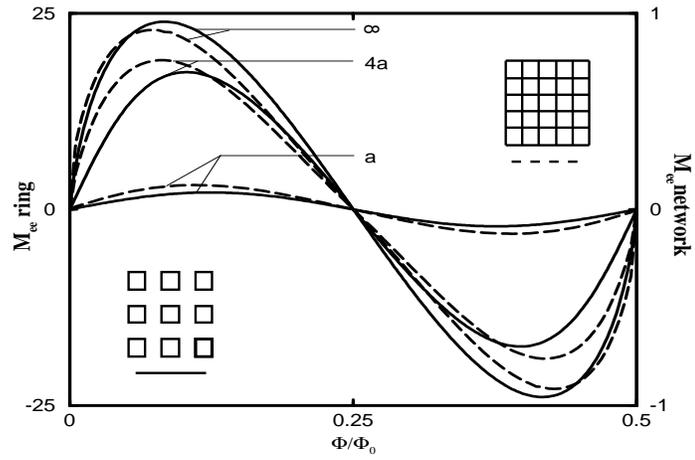}}
\caption{Average magnetization $\langle M_{ee} \rangle$ of a single ring (full lines)  and
magnetization
density of the infinite network (dashed lines), for $L_\phi=\infty$, $4 a$
and $a$ } \label{fig2}
\end{figure}

\end{document}